\documentclass[12pt,preprint]{aastex}
\begin{document}
%\submitted{}

\title{Model of ejection of matter from nonstationary dense stellar clusters
 and chaotic motion of gravitating shells.}

\author{M.V.Barkov}
\affil{Space Research Institute, 84/32 Profsoyuznaya Str , Moscow,
Russia, 117997;\\ barmv@sai.msu.ru}
\medskip
\author{V.A. Belinski}
\affil{National Institute of Nuclear physics (INFN) and
International Center of Relativistic Astrophysics (ICRA), Dip. di
Fisica - Universita` degli Studi di Roma "La Sapienza" P.le Aldo
Moro, 5 - 00185 Roma, Italy;\\ volodia@vxrmg9.icra.it}
\medskip
\author{G.S.Bisnovatyi-Kogan}
\affil{Space Research Institute, 84/32 Profsoyuznaya Str , Moscow,
Russia, 117997;\\ gkogan@mx.iki.rssi.ru}

\medskip

\begin{abstract}
 It is shown that during the motion of two
initially gravitationally  bound spherical shells, consisting of
point particles moving along ballistic trajectories, one of the
shell may be expelled to infinity at subrelativistic speed
$v_{exp}\leq 0.25 c$. The problem is solved in Newtonian gravity.
Motion of two intersecting shells in the case when they do not
runaway shows a chaotic behaviour. We hope that this toy and
oversimplyfied model can give nevertheless a qualitative idea on
the nature of the mechanism of matter outbursts from the dense
stellar clusters
\end{abstract}


\section{Introduction}.

Dynamical processes around supermassive black holes in quasars,
blazars and
 active galactic nuclei (AGN) are characterised by violent phenomena,
leading to formation of jets and other outbursts. Jet formation is
usually connected with the processes in the magnetized accretion
disks (Lovelace, 1976; Bisnovatyi-Kogan \& Blinnikov, 1976).
Formation of quasispherical outbursts, which are probably observed
in quasars with broad absorption lines, could be
 connected with another mechanism. Here we consider a possibility of a
shell outburst from a supermassive black holes (SBH) surrounded by
a dense massive stellar cluster, basing on a pure ballistic
interaction of gravitating shells oscillating around SBH.

Ballistic ejection may be responsible for appearance of stars and
observed SN in the intergalactic space, as well as the existence
of stars  between galaxies, suggested by many authors.

Investigation of spherical stellar clusters using shell
approximation was started by H\'enon (1964), and than have been
successfully applied for investigation of the stability H\'enon
(1973), violent relaxation and collapse (H\'enon 1964; Gott 1975),
leading to formation af a stationary cluster. Investigation of the
evolution of spherical stellar cluster with account of different
physical processes was done on the base of a shell model in the
classical serie of papers of L. Spitzer and his coauthors (Spitzer
\& Hart 1971a,b; Spitzer \& Thuan 1972; Spitzer \& Shapiro 1972;
Spitzer \& Chevalier 1973; Spitzer \& Shull 1975a,b; Spitzer \&
Mathiew 1980).

Numerical calculations of a collapse of stellar clusters in a
shell approximation (Yangurazova \& Bisnovatyi-Kogan 1984;
Bisnovatyi-Kogan \& Yangurazova 1987,1988) had shown, that even if
all shells are  initially gravitationally bound, after a number of
intersections some shells obtain sufficient energy to become
unbound, and to be thrown to the infinity. In the Newtonian
gravity the remnant is formed as a stationary stellar cluster, and
in general relativity SBH may be formed as a remnant.

Formation of outbursts due to the ballistic interactions may
happen as a result of intersection of shells, oscillating around
SBH. In the smooth cluster with or without SBH in the centre stars
evaporate, due to stellar encounters mainly with small kinetic
energy, and formation of rapidly expelling stars is about 100
times less probable due to predominance of gravitational
encounters with small momentum transfer Ambartsumian (1938). If
the cluster is strongly aggregated, consisting of few compact
pieces, the encounters between these pieces are quite different,
collision with large momentum transfer are becoming probable. In
this case gravitational interaction between compact pieces may
lead to the outburst with large velocity, and when such event
happens in the vicinity of SBH the velocity could become a
considerable part of the light velocity $c$.

Such kind of situation in principle may be produced as a result of
galactic collisions with close encounter of nuclei, when one of
the nuclei strips the matter from the companion in the form of a
collapsing shells. Interaction of such shell with the stellar
cluster may lead not only to collapse into SBH, but also to the
reverse phenomena: expelling of the shell with a speed, much
larger than the average speed of stars in the cluster. The shell
is not falling into SBH due to high angular momentum of its stars.

Even more important example of a quasi-spherical mass ejection is
a relativistic collapse of a spherical stellar system, which is
considered (Zeldovich \& Podurets, 1965; Lightman \& Shapiro 1978;
Ipser 1980; Rasio et al. 1989) as the main mechanism of a
formation of supermassive black holes in the galactic centers.
Approximation of such collapse by consideration of spherical
shells is the simplest approach, which reflects all important
features of such collapse (Gott 1975; Bisnovatyi-Kogan \&
Yangurazova 1987, 1988). Our consideration is related to the
motion of stars (shells) which remain outside the newly formed
supermassive black hole.

Here  we consider a simplified problem of a motion of two massive
spherical shells, each consisting of stars with the same specific
angular momentum and energies, around SBH.
    The aim of this paper is to give an elementary treatment of the
ballistic ejection, and to estimate the maximal efficiency of this
ejection. Note, that in a more complicated case of numerous shell
intersection this elementary act is a key process of the energy
exchange between stars and of matter ejection. This is also the
elementary process leading to the violent relaxation of the
cluster Lynden-Bell (1967), studied in the shell approximation by
Gott (1975) and Yangurazova \& Bisnovatyi-Kogan (1984).
Chaotization of the motion of two gravitating intersecting shells
appears as a by-prodiuct of our consideration. The appearence of
chaos in such a system, which is described by a set of simple
algebraic equations, is a rare example. For the oversimplified
case with a pure radial motion and reflecting inner boundary the
chaotic shell motion was found by Miller and Youngkins (1997).
Here we consider realistic shells consisting of stars moving along
eliptical trajectories, and their chaotic properties would be
examined in more details in another work.
 We find conditions at
which one of two shells is expelling to infinity taking energy
from another shell. We find a maximum of the velocity of the
outbursting shell as a function of the ratio of its mass $m$ to
the mass $M$ of SBH using only  Newtonian theory of the shell's
motion.

However we introduce cutoff  fixing the minimal radius of the
potentially outbursting shell $r_{m}$ on the level of few
$r_g={2GM}/{c^2}$. We show, that for equal masses of two shells
the expelling velocity reach the value $v_{max}\approx
0{.}3547v_p$ at $m/M=1.0$ , and $v_{max}\geq 0{.}3v_p$ was
obtained at the masses of shells $0.25\div1.5M$, where the
parabolic velocity of a shell in the point of a smallest distance
to the black hole $v_p$ may be of a considerable part of $c$.

In sections 2,3 we describe outburst effect and in  sections 4 we
present the evidence of the chaos in the system of intersecting
shells. The  exact solution of these problems in the context of
General Relativity will be presented in the subsequent paper.

\section{Two shells around SBH.}

Physically the nature of the ballistic ejection is based on the following
four subsequent events. The outer shell is accellerated
moving to the center in a strong gravitational field of a central body and
inner shell. Somewhere near the inner minimal radius of the trajectory shells
intersect. After that the former outer shell is deccellerating moving
from the center in the weaker gravitationsl field of only one central
body. The second intersection happens somwhere far from the center.
That may result in the situation when the total energy (negative)
of the initially gravitationally bound outer shell is becoming positive
as a result of two subsequent intersections with another shell.
The quantitive analysis of this procrss is done in this section.

Equation of motion of a shell with mass $m$ and total conserved
energy $E$ in the field of a central body with mass $M$ is
\begin{equation}
 \label{r1}
  E=\frac{mv^2}{2}-\frac{Gm(M+m/2)}{r}+\frac{J^2m}{2r^2},
\end{equation}
where $v=dr/dt$ is the radial velocity of the shell and
$J^2m/2r^2$ is the total kinetic energy of tangential motions of
all particles, the shell make-up from. The constant $J>0$ has that
interpretation that $Jm$ is the sum of the absolute values of the
angular momenta of all particles.The term $m/2$ in (\ref{r1}) is
due to the self-gravity of the shell.

Let's consider two shells with parameters $m_1,J_1$ and $m_2,J_2$
moving around SBH  with mass $M$. Let the shell "1"{} be initially
outer and the shell "2"{} be the inner one. Then equations of
motion are:
\begin{equation}
 \label{r2}
  E_{1(0)}=\frac{m_1v_{1(0)}^2}{2}-\frac{Gm_1(M+m_1/2+m_2)}{r} +
  \frac{J_1^2m_1}{2r^2},
\end{equation}
\begin{equation}
 \label{r3}
  E_{2(0)}=\frac{m_2v_{2(0)}^2}{2}-\frac{Gm_2(M+m_2/2)}{r} +
  \frac{J_2^2m_2}{2r^2}.
\end{equation}
By the index (0) we mark the initial evolution  stage before the
first intersection of the shells. Assume that both shells are
moving to the center and $v^2_{1(0)}>v^2_{2(0)}$. Such shells
intersect each other at a some radius $r=a_1$ and at some time
$t=t_1$ after wich the shell "1"{} becomes inner and shell "2"{}
-- outer. The equations of motion of the shells during this new
stage (designate it by the index (1)) are:
\begin{equation}
 \label{r4}
  E_{1(1)}=\frac{m_1v_{1(1)}^2}{2}-\frac{Gm_1(M+m_1/2)}{r} +
  \frac{J_1^2m_1}{2r^2},
\end{equation}
\begin{equation}
 \label{r5}
  E_{2(1)}=\frac{m_2v_{2(1)}^2}{2}-\frac{Gm_2(M+m_2/2+m_1)}{r} +
  \frac{J_2^2m_2}{2r^2}.
\end{equation}
The crucial point now are the matching conditions at the
intersection point $r=a_1,\quad t=t_1$ where from one can obtain
the initial data to the equations (\ref{r4}), (\ref{r5}) in order
to define uniquely the evolution during this new stage. In
Newtonian theory these conditions are:
 $$
 E_{1(0)} + E_{2(0)}= E_{1(1)} + E_{2(1)};
 $$
\begin{equation}
  \label{r6}
    v_{1(0)}(t_1)= v_{1(1)}(t_1);\qquad
    v_{2(0)}(t_1)= v_{2(1)}(t_1),
\end{equation}
i.e. the conservation of the total energy of the system and
continuity of the velocities through the intersection point.
Expressing $v_{1(0)}(t_1),\; v_{2(0)}(t_1)$ from eqs. (\ref{r2}),
(\ref{r3})  and $v_{1(1)}(t_1),\; v_{2(1)}(t_1)$ from eqs.
(\ref{r4}), (\ref{r5}) and equating them following conditions
(\ref{r6}) we get:
\begin{equation}
  \label{r7}
    E_{1(1)}= E_{1(0)}+\frac{Gm_1m_2}{a_1}; \qquad
   E_{2(1)}= E_{2(0)}-\frac{Gm_1m_2}{a_1}.
\end{equation}
At some point $r=a_2,\; t=t_2$ can happen the second intersection
after which the shell "1"{} again becomes the outer and shell
"2"{} becomes inner one. It is a simple task to write the equation
of motion during this third stage (we designate it by the index
(2)) and repeat the same procedure of matching velocities but now
at point $r=a_2,\; t=t_2$. The result is:
\begin{equation}
  \label{r8}
    E_{1(2)}= E_{1(1)}-\frac{Gm_1m_2}{a_2}=
    E_{1(0)}+Gm_1m_2\left(\frac1{a_1}-\frac1{a_2}\right),
\end{equation}
\begin{equation}
  \label{r9}
    E_{2(2)}= E_{2(1)}+\frac{Gm_1m_2}{a_2}=
    E_{2(0)}-Gm_1m_2\left(\frac1{a_1}-\frac1{a_2}\right).
\end{equation}
Let us describe the situation, when one shell is ejected to
infinity after intersection of to initially bound shells. We
consider a case when $a_2$ is larger then $a_1$ so, that the
second term  in (\ref{r8}) has larger absolute value, then the
first one, the first shell gains a positive energy and goes to
infinity. Both shells have initial negative energies $E_{1(0)}$
and $E_{2(0)}$, but with small enough absolute values. The first
shell takes the energy from the second one, which is becoming more
bound with larger absolute value of the negative energy
$E_{2(2)}$, according to (\ref{r8}).

\section{Numerical solution.}

Let's illustrate the foregoing scenario by an exact particular
example of two shells of equal masses. We choose parameters in the
following way:
\begin{equation}
 \label{r10}
 m_1=m_2=m,\qquad E_{1(0)}=E_{2(0)}=0,\qquad J_1<J_2,
\end{equation}
In fact such exact solution represents the first approximation to
the more general situation when $ E_{1(0)}$ and $E_{2(0)}$ are
nonzero (negative) but small in that sense that both modulus
$|E_{1(0)}|$ and $|E_{2(0)}|$ are much less than $Gm^2/a_1$.

We assume that initially both shells are moving towards the
center. It follows from eqs. (\ref{r2}), (\ref{r3}) that under
condition (\ref{r10}) such shells will intersect inescapably at
the point $r=a_1,\; t=t_1$. After the second intersection at
$r=a_2,\; t=t_2$ the shell "1"{} will be thrown to infinity with
expelling velocity $v_{1exp}$. It follows from (\ref{r8}) that
\begin{equation}
 \label{r12}
  v_{1exp}=\left. v_{1(2)}\right|_{r\rightarrow\infty}=
  \sqrt{2Gm\left(\frac1{a_1}-\frac1{a_2}\right)}.
\end{equation}
In order to construct a solution with maximal possible $v_{1exp}$
we consider a case when initially inner shell "2"{} reaches the
inner turning point at minimal possible radius $r=r_m$ (see
introduction), and intersects with the initially outer shell
after, during its outward motion.

To find the second intersection point we need to solve equation of
motion (\ref{r4}),(\ref{r5}) for shell's trajectories after the
first intersection. First, let us consider the  bound shell "2"{}.
We apply eq. (\ref{r3}) to the point $r=r_m$ which gives (see Fig.
1.)
\begin{equation}
 \label{r14}
  J^2_2=Gr_m(2M+m).
\end{equation}
%%%%%%%%%%%%%%%%%%%%%%%%%%%%%%%%%%%%%%%%% fig1
\begin{figure*}[t]
 \epsscale{1.0}
 \plotone{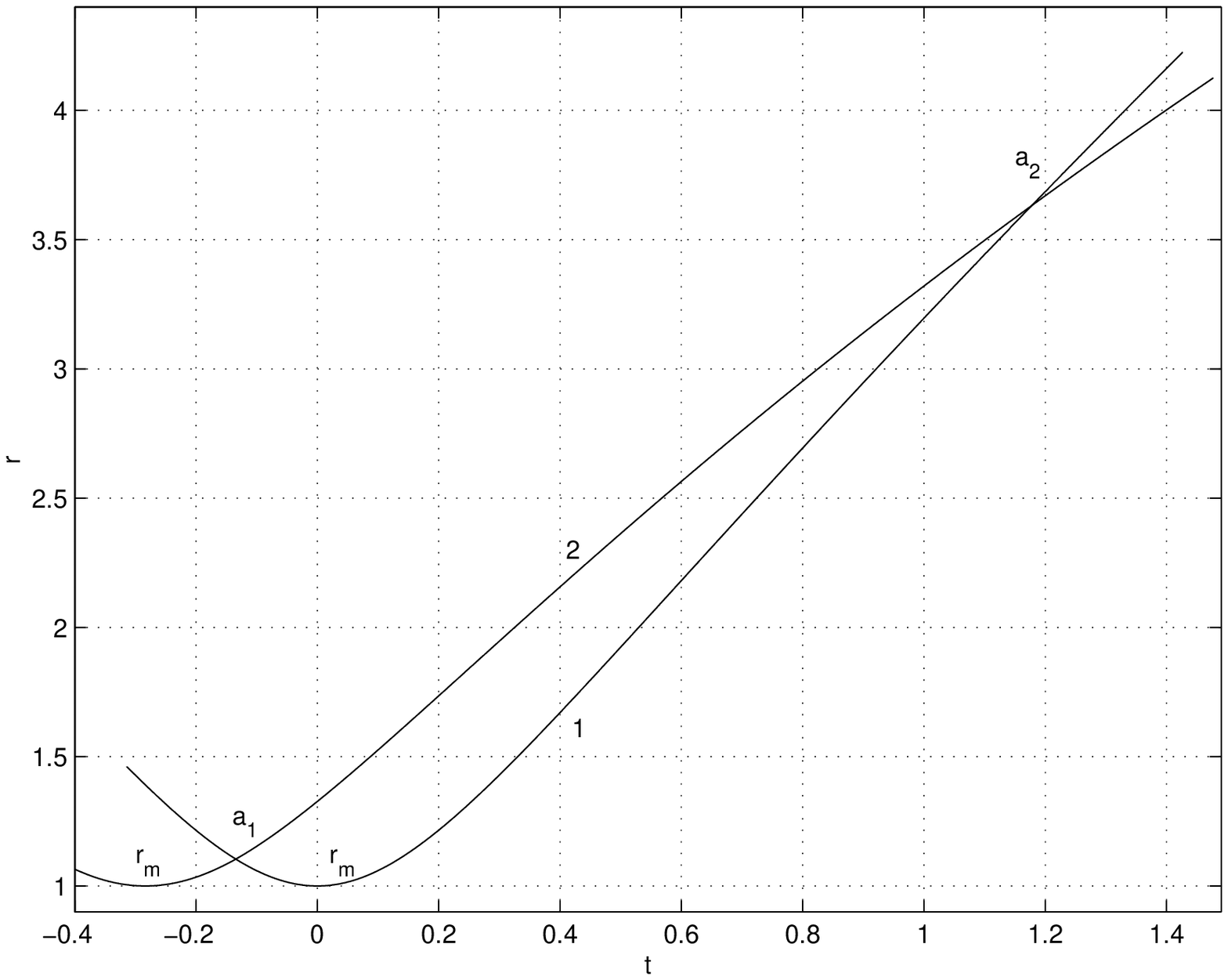}
 \caption{ Time dependence of radiuses r of two shells (in
units $r_m$) on time t (in units $r_m/v_p, \,\,
v_p=\sqrt{2G(M+3m/2)/r_m}$). The shells intersect at points
$a_1=1.1401 $, $a_2=3.6316 $. Here $m/M=0.1$, and after the second
intersection the first shell is running away with velocity at
infinity $v_{1exp}=0.23415 v_p$.}
 \label{Figure 1}
\end{figure*}
%%%%%%%%%%%%%%%%%%%%%%%%%%%%%%%%%%%%%%%%%%%%%%%%
Substituting this expression for $J^2_2$ into (\ref{r5}) and
taking into account (\ref{r7}), (\ref{r10}) we get a solution of
the equation (\ref{r5}) in the following parametric form:
\begin{equation}
 \label{nr13}
  \begin{array}{c}
 {\displaystyle  r(\eta)=b_{2(1)}(1-e_{2(1)}\cos \eta),} \\
 {\displaystyle t(\eta)=\sqrt{\frac{b_{2(1)}^3}{G(M+3m/2)}}
 (\eta - e_{2(1)}\sin \eta)-\delta_t.}
  \end{array}
\end{equation}
where the constants $b_{2(1)}$ and $e_{2(1)}$ are:
\begin{equation}
 \label{nr14}
  b_{2(1)}=\frac{M+3m/2}{2m}a_1, \qquad
  e_{2(1)}=\sqrt{1-\frac{4m(M+m/2)}{(M+3m/2)^2}\frac{r_m}{a_1}},
\end{equation}
on the choice of the constant $\delta_t$ see below.

Between the first and second intersection there exists the inner
turning point of the shell "1"{} (now inner shell). We take  that
additional restriction that the shell "1"{} reaches this turning
point also at the minimal possible radius $r=r_m$. It is easy to
show that a wide class of solutions with such restriction really
exists. At the point $r=r_m$ equation (\ref{r4}) gives:
\begin{equation}
 \label{nr15}
  J^2_1=Gr_m(2M+m)+2Gmr_m^2a_1^{-1}.
\end{equation}
We substitute this expression for $J^2_1$ into eq. (\ref{r4}) and
again with conditions (\ref{r7}) and (\ref{r10}) obtain the
following parametric form of the solution of this equation:
\begin{equation}
 \label{nr16}
  \begin{array}{c}
 {\displaystyle  r(\xi)=b_{1(1)}(e_{1(1)}\cosh \xi-1),} \\
 {\displaystyle t(\xi)=\sqrt{\frac{b_{1(1)}^3}{G(M+m/2)}}
 (e_{1(1)}\sinh \xi-\xi),}
  \end{array}
\end{equation}
where the constant $b_{1(1)}$ and $e_{1(1)}$ are:
\begin{equation}
 \label{nr17}
  b_{1(1)}=\frac{M+m/2}{2m}a_1, \quad
  e_{1(1)}=1+\frac{2m}{M+m/2}\frac{r_m}{a_1},
\end{equation}
The first intersection point $r=a_1,\; t=t_1$ corresponds to some
parameters $\eta=\eta_1,\; \xi=\xi_1$. At this point the time
$t(\eta_1)$ calculated from (\ref{nr13}) should coincide with time
$t(\xi_1)$ calculated from (\ref{nr16}). This fix the constant
$\delta_t$:
\begin{equation}
 \label{nr18}
 \begin{array}{c}
 {\displaystyle  \delta_t= \sqrt{\frac{b_{2(1)}^3}{G(M+3m/2)}}
 (\eta_1 - e_{2(1)}\sin \eta_1)-} \\
 {\displaystyle \sqrt{\frac{b_{1(1)}^3}{G(M+m/2)}}
 (e_{1(1)}\sinh \xi_1-\xi_1 ).}
 \end{array}
\end{equation}
Also the values of $r(\eta_1)$ from (\ref{nr13}) and $r(\xi_1)$
from (\ref{nr16}) should be the same and equal to $a_1$. From this
follows parameters $\eta_1$ and $\xi_1$:
\begin{equation}
 \label{nr19}
   \cosh\xi_1=\frac{b_{1(1)}+a_1}{b_{1(1)}e_{1(1)}}
\end{equation}
\begin{equation}
 \label{nr20}
  \cos\eta_1=\frac{b_{2(1)}-a_1}{b_{2(1)}e_{2(1)}}
\end{equation}
The second intersection point $r=a_2$ corresponds to parameters
$\eta=\eta_2,\; \xi=\xi_2$. The application of the analogous
coincidence conditions to the equations (\ref{nr13}) and
(\ref{nr16}) at this second point gives:
\begin{equation}
 \label{nr21}
  a_2=b_{1(1)}(e_{1(1)}\cosh \xi_2-1)
\end{equation}
\begin{equation}
 \label{nr22}
  a_2=b_{2(1)}(1-e_{2(1)}\cos \eta_2)
\end{equation}
\begin{equation}
 \label{nr23}
  \begin{array}{c}
 {\displaystyle  \sqrt{\frac{b_{1(1)}^3}{G(M+m/2)}}
 (e_{1(1)}\sinh \xi_2-\xi_2 - e_{1(1)}\sinh \xi_1+\xi_1)=} \\
 {\displaystyle \sqrt{\frac{b_{2(1)}^3}{G(M+3m/2)}}
 (\eta_2 - e_{2(1)}\sin \eta_2-\eta_1 + e_{2(1)}\sin \eta_1).}
  \end{array}
\end{equation}
From (\ref{nr14}),(\ref{nr17}) and (\ref{nr19})-(\ref{nr23}) it is
easy to see that all these relations determine $a_2/a_1$ as a
function of $a_1/r_m$ and $m/M$.

We introduce now the "parabolic" velocity $v_p$ of the any outer
shell at the point of its minimal distance to the center $r=r_m$
as:
\begin{equation}
 \label{r25}
 v_p=\sqrt{\frac{2G(M+3m/2)}{r_m}}.
\end{equation}
Than we have
\begin{equation}
 \label{r26}
 \frac{v_{1exp}}{v_p}=\sqrt{\frac{m}{M+3m/2}
 \left(\frac{r_m}{a_1}-\frac{r_m}{a_2}\right)}.
\end{equation}
Consequently this ratio is also a function of $a_1/r_m$ and $m/M$.
We find numerically the value of $a_1/r_m$ maximizing the ratio
$v_{1exp}/v_p$ which value is a function of $m/M$ only.
Dependencies for $v_{1exp}/v_p$, $a_2/a_1$, and $a_1/r_m$ as
functions of the parameter $m/M$ for these maximizing solutions
are given in Fig. 2-4.

%%%%%%%%%%%%%%%%%%%%%%%%%%%%%%%%%%%%%%%%% fig2
\begin{figure*}[t]
 \epsscale{1.0}
 \plotone{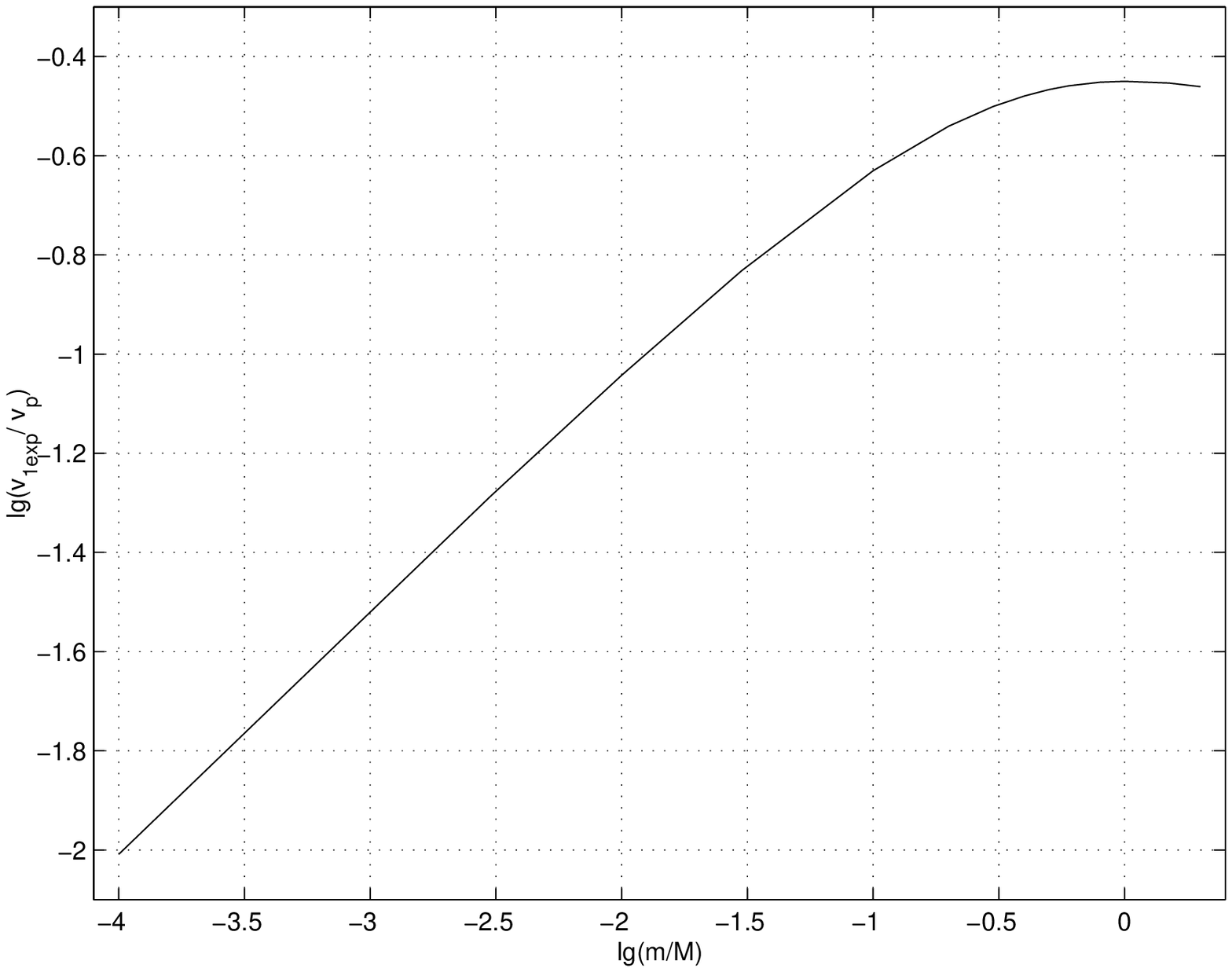}
 \caption{ Dependence of the maximum runaway velocity
$v_{1exp}$ in units $v_p$ on the ratio of the  shell mass to the
mass of central body $m/M$. The maximum value of $v_{1exp}=0.3547
$ corresponds to $m/M=1.0 $. }
 \label{Figure 2}
\end{figure*}
%%%%%%%%%%%%%%%%%%%%%%%%%%%%%%%%%%%%%%%%%%%%%%%%

%%%%%%%%%%%%%%%%%%%%%%%%%%%%%%%%%%%%%%%%% fig3
\begin{figure*}[t]
 \epsscale{1.0}
 \plotone{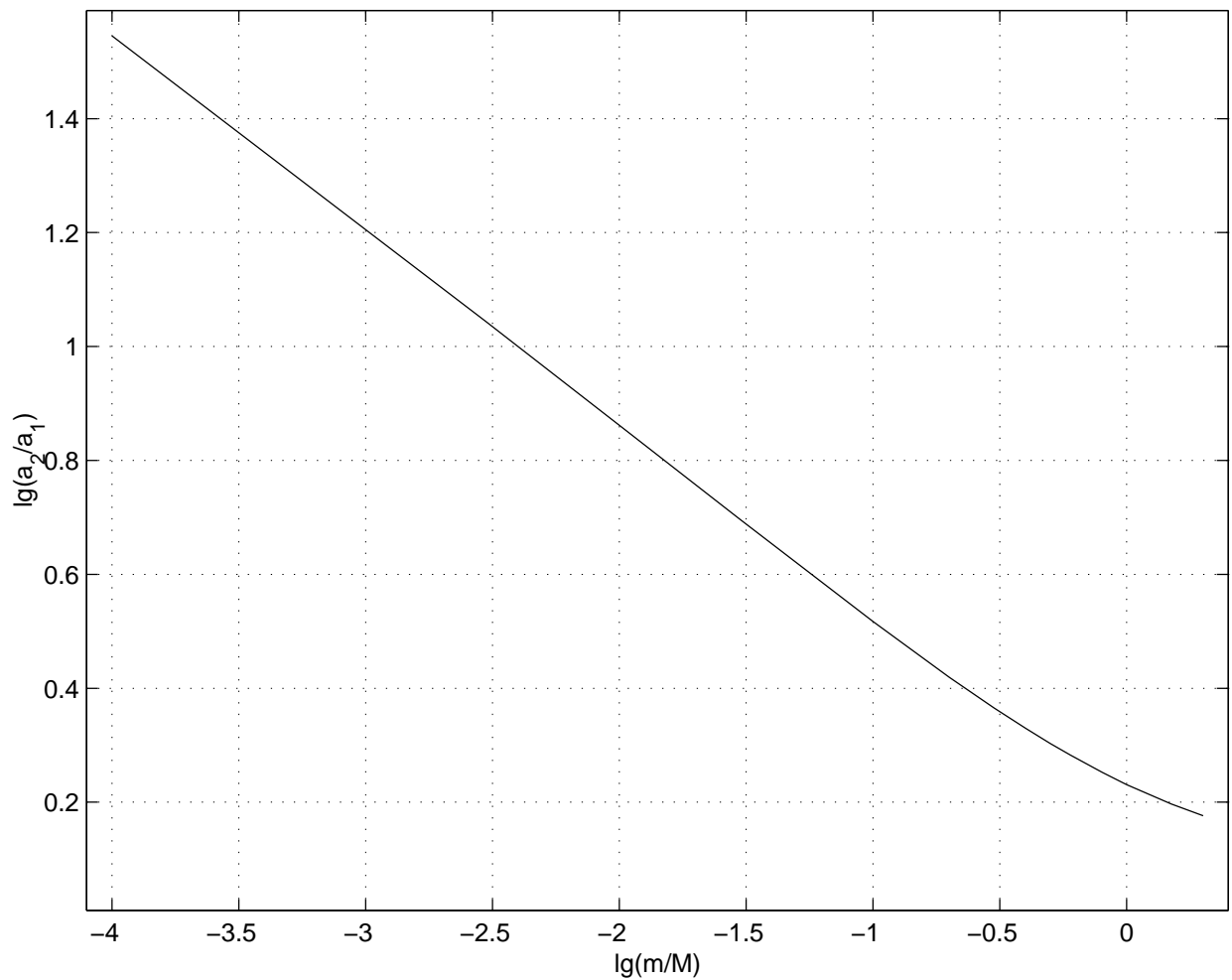}
 \caption{  Dependence of the ratio of intersection radiuses
$a_2/a_1$ on $m/M$ for the solutions with maximum runaway
velocities.}
 \label{Figure 3}
\end{figure*}
%%%%%%%%%%%%%%%%%%%%%%%%%%%%%%%%%%%%%%%%%%%%%%%%

%%%%%%%%%%%%%%%%%%%%%%%%%%%%%%%%%%%%%%%%% fig4
\begin{figure*}[t]
 \epsscale{1.0}
 \plotone{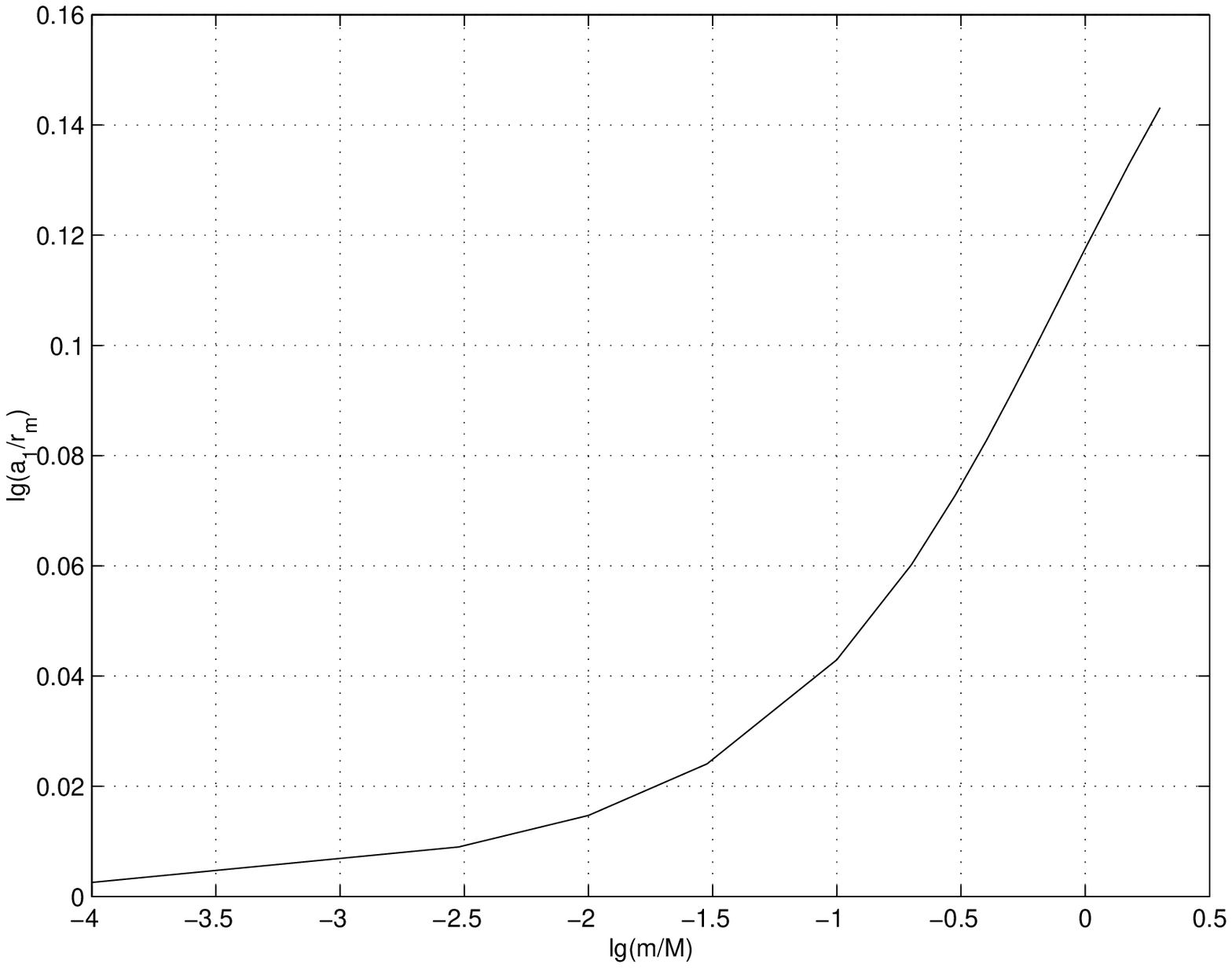}
 \caption{  The value  $a_1/r_m $ (the ratio
of the radius of the first intersection $a_1$ to the minimal
radius of the approach to the central body $r_m$)
as a function of $m/M$ for solutions
with maximum runaway velocities.}
 \label{Figure 4}
\end{figure*}
%%%%%%%%%%%%%%%%%%%%%%%%%%%%%%%%%%%%%%%%%%%%%%%%

The numerical calculations (illustrated by the Fig. 2) shows that
the runaway velocity $v_{1exp}$ reaches its maximal possible value
at $m/M\approx1.0$ and it is $v_{1exp}\approx0.3547v_p$. If we
consider the shells around SBH then the minimal radius $r_m$ of
the shell "1"{} orbit cannot be less then the two gravitational
radiuses  $2r_g=4G(M+3m/2)/c^2$. In the extreme case when $r_m=
2r_g$ we have $v_p\sim c/\sqrt{2}$ and for the maximal possible
runaway velocity in the family of solutions, characterized by
(\ref{r14}), (\ref{nr15}) we get $v_{1exp}\sim0.25c$.

\section{Chaos in the shell motion.}

The first evidence that the motion of two intersecting shells can
show chaotic character was given by B.N. Miller \& V.P. Youngkins
(1997). They investigated the special case when the central body
is absent and particles consisting the shells are moving only in
radial direction (in our notation $M=0$ and $J_1=J_2=0$). This
situation, however, cannot model astrophysical cluster with
massive nuclei, and also the problem of the influence of central
Newtonian singularity arise which need some additional care. In
any case a study of more physically realistic models with nonzero
$M, J_1$ and $J_2$ from the point of view of possible chaotic
behaviour represents essential interest. We report here some
results for such more general two shells model which was
investigated in the previous sections but again for the shells
with equal masses. We consider now only oscillatory regime of
motion without any runaway effects.

The shell motion  in the Newton gravitational  field  is
completely regular, but at presence of intersections the picture
changes qualitatively. The shell intersections result to chaos in
their motions. Character of chaos depends, mainly, on  mass ratio
of a shell and a  central body. At a small shell mass the motion
of the shells occurs basically in the field of the central body,
and after one intersection there is a little change in a
trajectory of each shell. Consider a case when angular momentum
parameters $J$ and energies of shells are close to each other, and
as a result of two intersections energy is transferred to the
shell "1"{}. Since shell parameters changes are not large, the
mutual positions and speeds of shells after intersections do not
change strongly. Then at the next intersection the part of energy
will again be transferred to the shell "1". After a large number
of intersections, however, the  mutual positions  changes
essentially. Then the energy  starts to flow  to the shell "2",
consequently we observe beating in oscillation of shells. In the
case when the masses of the intersecting shells  are large enough,
the energy is transferred very intensively between them. It
changes trajectories of shells essentially so that the next
intersection does not resemble at all the previous one. In this
case a shell's motion at once gets properties of a randomness.

%%%%%%%%%%%%%%%%%%%%%%%%%%%%%%%%%%%%%%%%% fig5
\begin{figure*}[t]
 \epsscale{1.0}
 \plotone{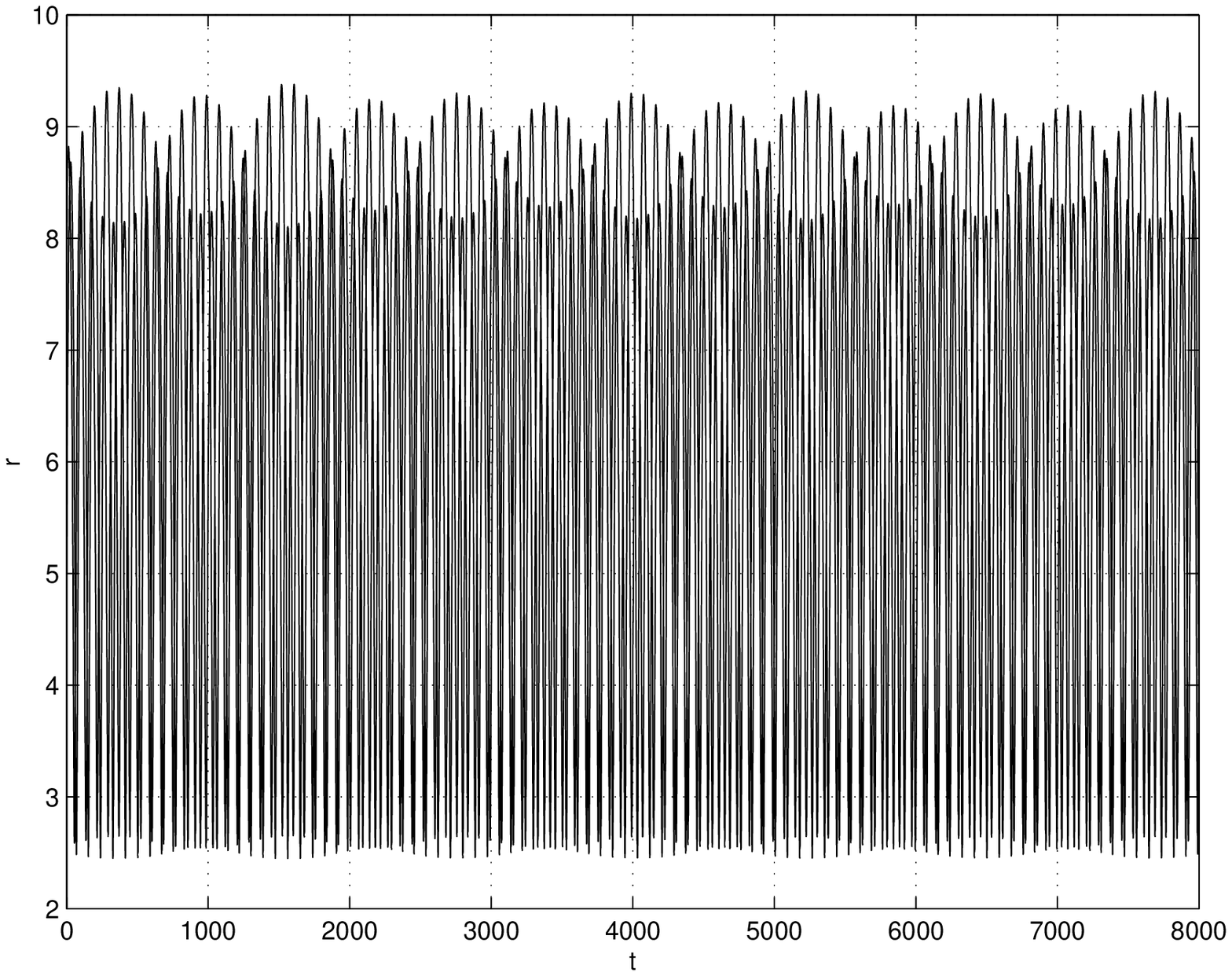}
 \caption{ Shell oscillations with mass ratio $m/M=0.015$.}
 \label{Figure 5}
\end{figure*}
%%%%%%%%%%%%%%%%%%%%%%%%%%%%%%%%%%%%%%%%%%%%%%%%

Shell oscillations with mass ratio $m/M=0.015 $ are presented in
Fig.5. In the case of massive shells the exchange of energy
between shells occurs more intensively. As a result we have the
obviously expressed chaotic behaviour of shells. Chaotic shell
oscillations with $m/M=0.08 $ are presented in Fig.6. In Fig.6 a
full randomness of behaviour of the shells is seen obviously.

%%%%%%%%%%%%%%%%%%%%%%%%%%%%%%%%%%%%%%%%% fig6
\begin{figure*}[t]
 \epsscale{1.0}
 \plotone{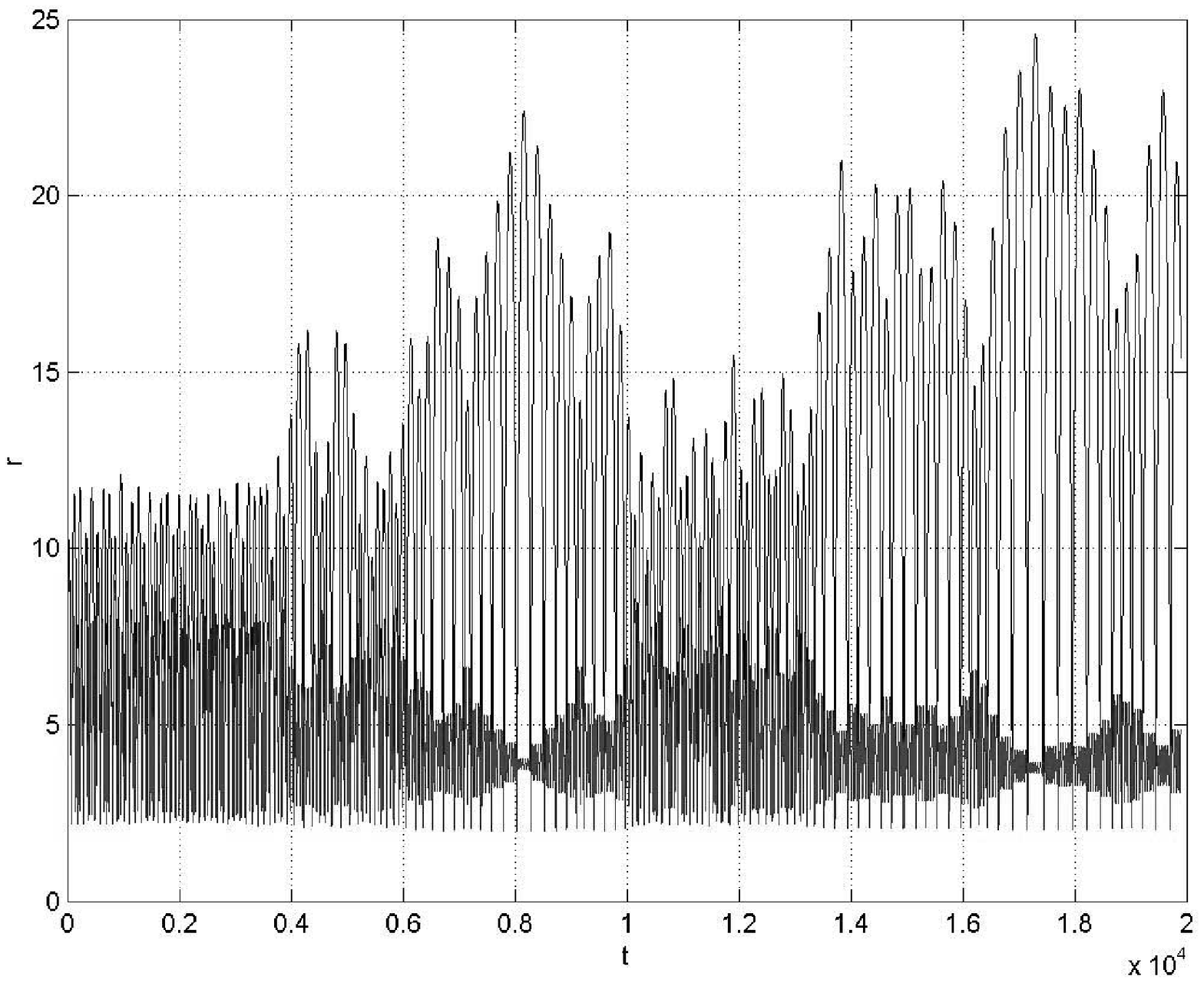}
 \caption{  Chaotic shell oscillations with mass ratio
$m/M=0.08$.}
 \label{Figure 6}
\end{figure*}
%%%%%%%%%%%%%%%%%%%%%%%%%%%%%%%%%%%%%%%%%%%%%%%%

Let us note, that when the exchange of energy between shells is
sufficiently large the strong randomness appears. Shells may
exchange their positions during the motion. The external shell may
become inner one and vice versa. One of the shells may give up
practically all its energy of the orbital motion, which it may
loose at constant angular momentum, i.e. the orbits of its
particles become almost circular. It may be seen in Figs.6,7 at those
moments when the orbit of the inner shell fills a narrow band by
itself.

Let us note that in our calculations the motion  of the shells is
very sensitive to initial parameters, what gives an additional
evidence that intersecting shells represent a really chaotic
system. When we change only precision of integrator from $10^{-6}$
 to $3\times 10^{-7}$, after several intersection
motion of shells becomes absolutely different from its initial
behaviour in Fig.6 at the same initial conditions. If we change the angular
momentum parameter of one of shell from 1.40 to 1.41, then in the
same way as in previous case, motion of shells becomes absolutely
different after several intersection. We can make statement that
the behaviour of shells is unstable and is strongly affected by
any perturbation in initial parameters of shell motion.

%%%%%%%%%%%%%%%%%%%%%%%%%%%%%%%%%%%%%%%%% fig7
\begin{figure*}[t]
 \epsscale{1.0}
 \plotone{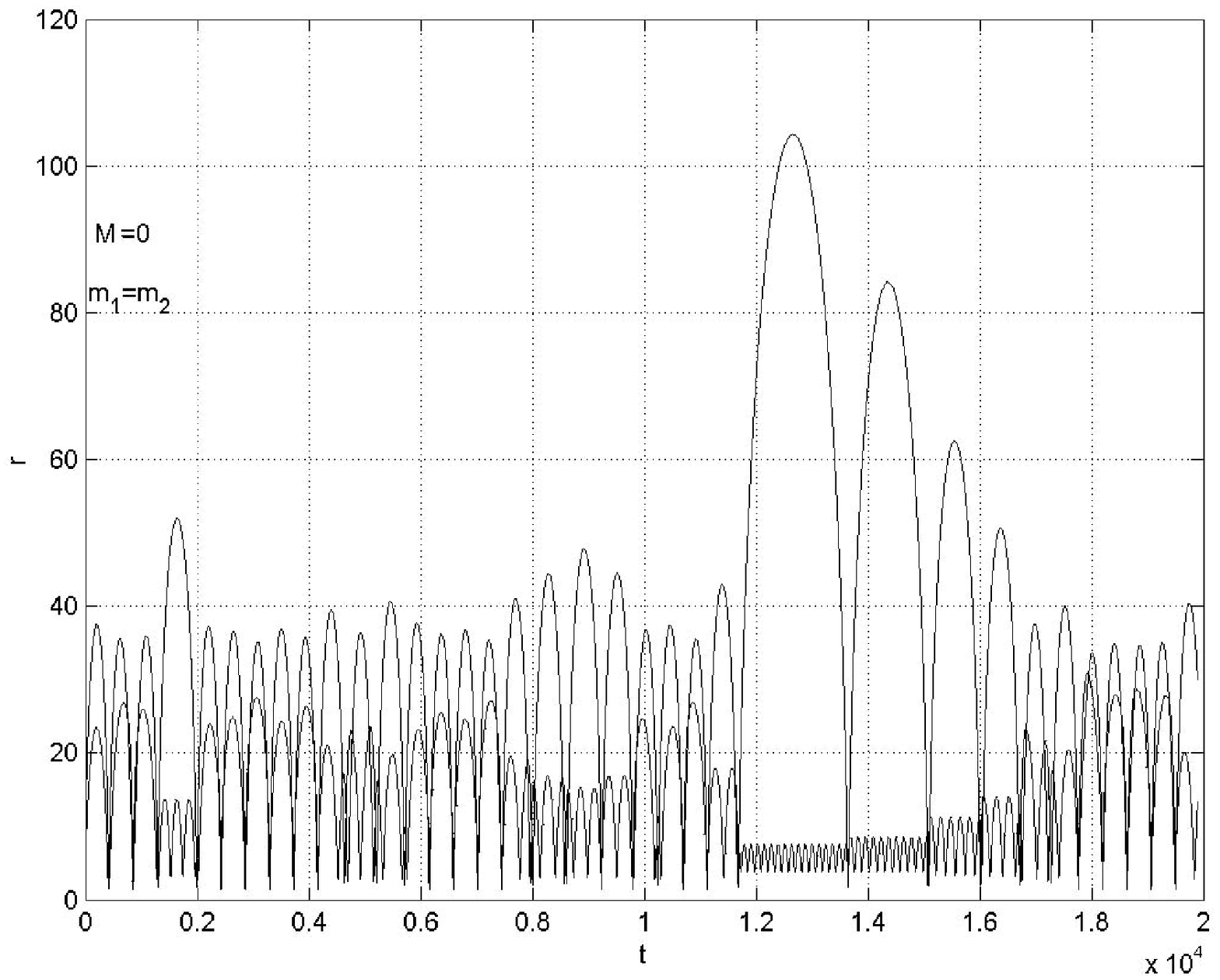}
 \caption{  The chaotic motion $r_{1,2}(t)$ of two
selfgravitating shells in their own gravitational field without a
central mass.}
 \label{Figure 7}
\end{figure*}
%%%%%%%%%%%%%%%%%%%%%%%%%%%%%%%%%%%%%%%%%%%%%%%%

The example of a chaotic behaviour of two intersecting
selfgraviting shells, moving in their own gravitational field,
without a central mass, is represented in Fig. 7.

Chaotic behaviour of the intersecting selfgravitating shells in
the Newton gravity is a rare example of a chaotic dynamical
system, described fully analytically by algebraic relations.

\section{Discussion}

 We have shown that pure gravitational interaction of two
spherical shells and central object may lead to considerable
energy exchange and ejection of one of the shell with a
subrelativistic speed $\sim0.25c$. Motion of stars around SBH may
lead to similar effects due to their interaction, and we may
expect formation of very rapidly moving stars, which could
overcome the galactic gravity and runaway into the intergalactic
space. In this way very extended galactic halo may be formed.
Ballistic mechanism of energy exchange between gravitating
particles may be important in the evolution of spatial
distribution of cold dark matter, when gravitational instabilities
are developing in masses, much less than galactic ones, and
gravitationally bound massive objects, consisting mainly from the
dark matter interacting only due to its gravitation, may be
formed. Interesting examples of ballistic interaction of
satellites with planets and stars, their acceleration and
ballistic control are presented in the book of V.V. Beletskyi
(1977).

\section*{Acknowledgement}

The work of G.S.B.-K. and M.V.B. was partly supported by RFBR
grant 99-02-18180, and INTAS grant 00-491. We are grateful to the
referee, Prof. Seppo Mikkola for useful comments.

%\newpage

\newpage

\figurename\,  1. Time dependence of radiuses r of two shells (in
units $r_m$) on time t (in units $r_m/v_p, \,\,
v_p=\sqrt{2G(M+3m/2)/r_m}$). The shells intersect at points
$a_1=1.1401 $, $a_2=3.6316 $. Here $m/M=0.1$, and after the second
intersection the first shell is running away with velocity at
infinity $v_{1exp}=0.23415 v_p$.

\figurename\,  2. Dependence of the maximum runaway velocity
$v_{1exp}$ in units $v_p$ on the ratio of the  shell mass to the
mass of central body $m/M$. The maximum value of $v_{1exp}=0.3547
$ corresponds to $m/M=1.0 $.

\figurename \, 3. Dependence of the ratio of intersection radiuses
$a_2/a_1$ on $m/M$ for the solutions with maximum runaway
velocities.

\figurename\,  4. The value  $a_1/r_m $  (the ratio
of the radius of the first intersection $a_1$ to the minimal
radius of the approach to the central body $r_m$)
as a function of $m/M$ for solutions
with maximum runaway velocities.

\figurename\,  5.  Shell oscillations with mass ratio $m/M=0.015$.

\figurename\,  6. The chaotic motion $r_{1,2}(t)$ of shells with
masses $m/M=0.08$.

\figurename\,  7. The chaotic motion $r_{1,2}(t)$ of two
selfgravitating shells in their own gravitational field without a
central mass.

\end{document}